\documentclass[10pt,conference]{IEEEtran} 
\IEEEoverridecommandlockouts
\usepackage{cite}
\usepackage{url}
\usepackage{amsmath,amssymb,amsfonts}
\usepackage{algorithmic}
\usepackage{graphicx}
\usepackage{tcolorbox}
\usepackage{textcomp}
\usepackage{xcolor}
\usepackage{xspace}
\usepackage{cleveref}
\usepackage{booktabs}
\usepackage{subcaption}
\usepackage{multirow}
\usepackage{multicol}
\usepackage{subcaption}

\def\BibTeX{{\rm B\kern-.05em{\sc i\kern-.025em b}\kern-.08em
    T\kern-.1667em\lower.7ex\hbox{E}\kern-.125emX}}
\begin{document}

\title{DANDI: Diffusion as Normative 
Distribution for Deep Neural Network Input
\thanks{This work has been supported by the National Research Foundation of Korea (NRF) funded by the Korean government MSIT (RS-2023-00208998), and the Engineering Research Center Program funded by the Korean Government MSIT (RS-2021-NR060080).}}

\author{\IEEEauthorblockN{Somin Kim}
\IEEEauthorblockA{\textit{School of Computing, KAIST} \\
Daejeon, Republic of Korea \\
somin.kim@kaist.ac.kr}
\and
\IEEEauthorblockN{Shin Yoo}
\IEEEauthorblockA{\textit{School of Computing, KAIST} \\
Daejeon, Republic of Korea \\
shin.yoo@kaist.ac.kr}
}

\maketitle

\newcommand{\longname}{\textsc{[LONGNAME]}\xspace}
\newcommand{\name}{\textsc{DANDI}\xspace}
\newcommand{\fixme}[1]{{\color{red}\textbf{FIXME:}(#1)}}
\newcommand{\fixed}[1]{{\color{blue}\textbf{FIXED:}(#1)}}
\newcommand{\addcite}[1]{{\color{orange}\textbf{addcite:}#1}}

\begin{abstract}
Surprise Adequacy (SA) has been widely studied as a test adequacy metric that can effectively guide software engineers towards inputs that are more likely to reveal unexpected behaviour of Deep Neural Networks (DNNs). Intuitively, SA is an out-of-distribution metric that quantifies the dissimilarity between the given input and the training data: if a new input is very different from those seen during training, the DNN is more likely to behave unexpectedly against the input. While SA has been widely adopted as a test prioritization method, its major weakness is the fact that the computation of the metric requires access to the training dataset, which is often not allowed in real-world use cases. We present \name, a technique that generates a surrogate input distribution using Stable Diffusion to compute SA values without requiring the original training data. An empirical evaluation of \name applied to image classifiers for CIFAR10 and ImageNet-1K shows that SA values computed against synthetic data are highly correlated with the values computed against the training data, with Spearman Rank correlation value of 0.852 for ImageNet-1K and 0.881 for CIFAR-10. Further, we show that SA value computed by \name achieves can prioritize inputs as effectively as those computed using the training data, when testing DNN models mutated by DeepMutation. We believe that \name can significantly improve the usability of SA for practical DNN testing.
\end{abstract}


\section{Introduction}

Deep Neural Networks (DNNs) have been rapidly adopted into safety critical systems such as autonomous driving vehicles and medical imaging devices, resulting in urgent needs to test these systems. While DNNs suffer from a range of faults~\cite{Humbatova2020kt}, testing of DNNs remains a challenge as, typically, the test oracle can only be provided by humans and are extremely expensive. Consequently, various test adequacy metrics for test inputs have been proposed, so that the tester can prioritise test inputs for DNNs according to their likelihood to reveal incorrect behaviour~\cite{Pei2017qy,Tian2018aa,Feng2020mw,Kim2019aa}.

Surprise Adequacy (SA)~\cite{Kim2019aa,Kim2022hg} is a widely studied test adequacy metric. Intuitively, SA is an out-of-distribution-ness measure: it quantifies the difference between the current input and the data seen during the training. The measurement is made via Activation Traces (ATs), i.e., the activation values of all neurons in a chosen layer of DNN during the forward inference of a specific given input. ATs can be thought to capture the internal behaviour of the model against the input. If the AT produced by the current input is similar to those produced by the training data, the model is likely to perform well with the input; if the AT produced by the current input is not similar to those from the training data, the model is likely to perform unexpectedly. 

SA has been shown to be effective against image classifiers~\cite{Kim2019aa,Kim2022hg}, semantic segmentation models for autonomous driving~\cite{Kim2020zg}, as well as RNN models that takes textual inputs~\cite{Kim2020aa,Kim2021ct}. SA has also been used as guidance to synthesize test input images near the class boundaries~\cite{Kang2020aa,Kang2024ac}. Despite its effectiveness, SA has also been criticized for one major limitation~\cite{Yuan2023aa}, which is that its computation requires the access to the training data. There may be many scenarios that do not allow such access: the model may be pre-trained, or the training data may include private or copyrighted material. 

We propose \name (Diffusion as Normative Distribution for DNN Input), a technique that allows the computation of SA without the access to the training dataset. \name is based on two core assumptions. First, we note that, to compute SA values, it is critical to synthesise the distribution of \emph{normative} inputs, i.e., those that can represent the majority of the specific input class under consideration. For example, an image classifier for fruits would be trained using a large number of good quality apples, unless the classifier is to be used to pick out apples that have gone bad. Second, the recent advances in generative models mean that it is possible for them not only to produce synthetic test inputs to test models that are trained and used with real inputs, but also to explicitly specify the class of inputs to synthesize. For example, to test a fruit classification model that is part of grocery store checkout machine, we can synthesise inputs of good looking apples, whereas to test a model that is part of a production line in apple jam factory, we can also synthesise worm-eaten apples. For both use cases, the synthetic inputs would be realistic enough to be used as inputs to the model under test.

The paper evaluates the use of generative models as a surrogate input distribution to compute SA. While the paper instantiates \name for image classifiers, using Stable Diffusion as the generative model, we believe the core assumptions described above apply to other modalities such as natural language. We first show that distributions of SA values produced using the real training data and the synthetic surrogate data are statistically indistinguishable, using ImageNet1K and CIFAR10 datasets. Subsequently, we also show that, when prioritising test inputs, there exists a high correlation between the order produced by SA computed with training data, and SA computed with synthetic data. Finally, we show that \name can successfully prioritise inputs to kill DNN model mutants produced by DeepMutation~\cite{ma2018deepmutation}.

The remainder of the paper is structured as follows. Section \ref{sec:background} provides the academic backdrop to our paper and motivates our approach. Section \ref{sec:approach} describes in particular how we used a diffusion model to generate images for DNN testing. Section \ref{sec:rqs} details our specific research questions and the experimental setup that we used to obtain our results. In Section \ref{sec:results}, the results of our analysis are provided; with threats to validity discussed in Section \ref{sec:validity}. We discuss potential future directions and conclude in Section \ref{sec:conclusion}.

\section{Background}
\label{sec:background}

\subsection{Surprise Adequacy}

Surprise Adequacy (SA) is a widely studied DNN test adequacy metric that essentialy measures the similarity between the given input and the data encountered during the model's training~\cite{Kim2019aa}. The intuition is that the model is more likely to perform correctly when given inputs similar to the training data, i.e., less surprising. Conversely, more surprising inputs, i.e., those that differ significantly from the training data, are less likely to be handled accurately by the model. Among the multiple ways of calculating Surprise Adequacy, this paper adopts Likelihood-based Surprise Adequacy (LSA). 
LSA first represents each input using its Activation Trace (AT) vector, which is the output of a neural layer captured while processing that input. By gathering Activation Traces from all training inputs, one captures the model’s internal representation of the training data. Subsequently, LSA applies Kernel Density Estimation (KDE) using the Gaussian kernel function. When a new input is given, its LSA value is computed as the negative logarithm of this density. A low LSA value indicates that the input is similar to the training data, suggesting that the model is likely to perform accurately. Conversely, a high LSA value signifies that the input is different from the training data, implying that the model may be less reliable in handling it.

SA has been applied to image classification~\cite{Kim2019aa,Kim2022hg}, object segmentation for autonomous driving~\cite{Kim2020zg}, question and answering~\cite{Kim2020aa} as well as text classification~\cite{Kim2021ct}. However, its major weakness is that one needs the access to training data to compute SA values, which may not be the case in real-world scenario. This paper aims to address this limitation.

\subsection{Mutation Testing for DNNs}
DNN models are typically evaluated using test datasets, therefore the quality of these datasets is crucial; inadequate test sets can result in models that appear accurate but lack generality and robustness. 

Mutation testing is a traditional software technique that injects artificial faults to evaluate the fault-detection capabilities of test suites~\cite{jia2010analysis}. However, conventional mutation operators are not directly applicable to DL systems due to fundamental differences; traditional software operates with explicit logic and deterministic control flow, while DL models are data-driven and rely on learned representations from training datasets. Tools like DeepMutation and DeepCRIME~\cite{ma2018deepmutation,Humbatova2021aa} address this by proposing DNN-specific mutation operators.

DeepMutation~\cite{ma2018deepmutation} provides source-level and model-level mutation operators. Source-level operators modify training data or model structure before training, requiring retraining. In contrast, model-level operators adjust a trained model's weights and biases, thus avoiding retraining and offering higher efficiency. Due to the high cost of retraining on our datasets, we employ model-level operators, specifically Gaussian Fuzzing (GF), which introduces Gaussian noise into model weights by scaling them. The operator is defined as:

\[
\text{GF}(W, \rho, \sigma) =  w \cdot (1 + \epsilon)
\]

where the weights \( w \) to be mutated are sampled uniformly from \( W \) with probability \( \rho \), and \( \epsilon \) is sampled from \( \mathcal{N}(0, \sigma^2) \), altering weights by approximately \( 100 \times \sigma\% \) (default \( \sigma = 0.5 \)). In DeepMutation, a mutant model is considered \textit{killed} if it misclassifies a test data point that the original model classifies correctly. This criterion assesses the test set's effectiveness in detecting introduced faults. DeepCRIME~\cite{Humbatova2021aa} introduces a statistical killing criterion that accounts for the inherent randomness in model training and mutant generation. Unlike DeepMutation, which defines killing criteria based on single-instance mutants, DeepCRIME leverages multiple instances (20 by default) for each mutant for the same mutation operator. This approach allows for a statistical definition of mutant killing. A mutant model is \textit{killed} if, against a test set, statistical analysis identifies a significant difference with a meaningful effect size in output quality metrics, such as accuracy, between the original and mutant models. 

To assess the effectiveness of our approach, we employ mutation testing to determine how well \name-based prioritized input set kills mutants, aiming to verify the dataset's quality and demonstrate the efficacy of \name-based prioritization.

\begin{figure}[t!]
    \centering
    \includegraphics[width=\columnwidth]{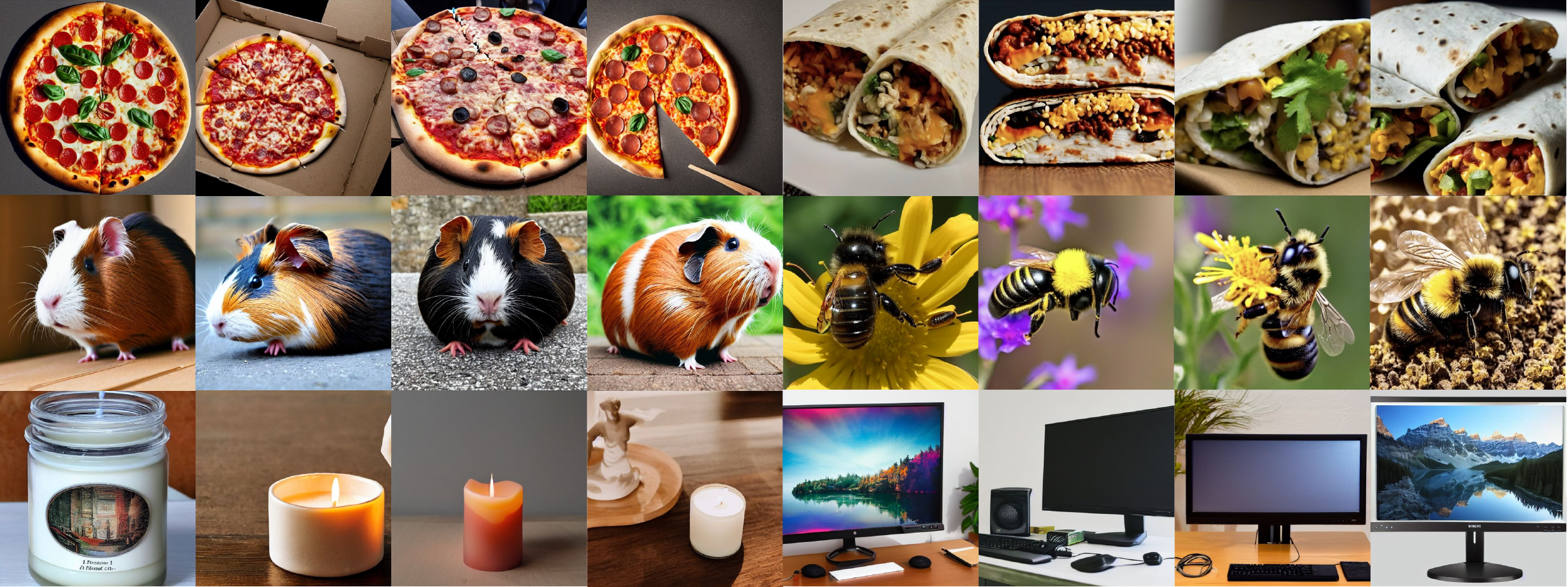}
    \caption{Examples of generated images with \name for classes pizza, burrito, guinea pig, bee, candle and monitor.}
    \label{fig:examples}
\end{figure}

\begin{figure*}[ht]
    \centering
    \begin{subfigure}[b]{0.49\textwidth}
    \centering
    \includegraphics[scale=0.3]{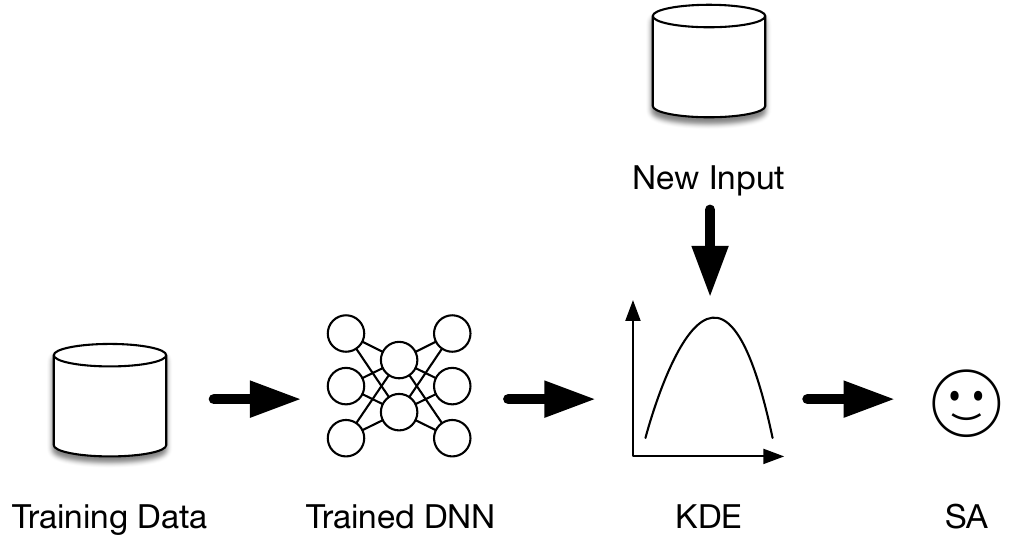}
    \caption{Original Surprise Adequacy Workflow\label{fig:sa}}
    \end{subfigure}
    \hfill
    \begin{subfigure}[b]{0.49\textwidth}
    \centering
    \includegraphics[scale=0.3]{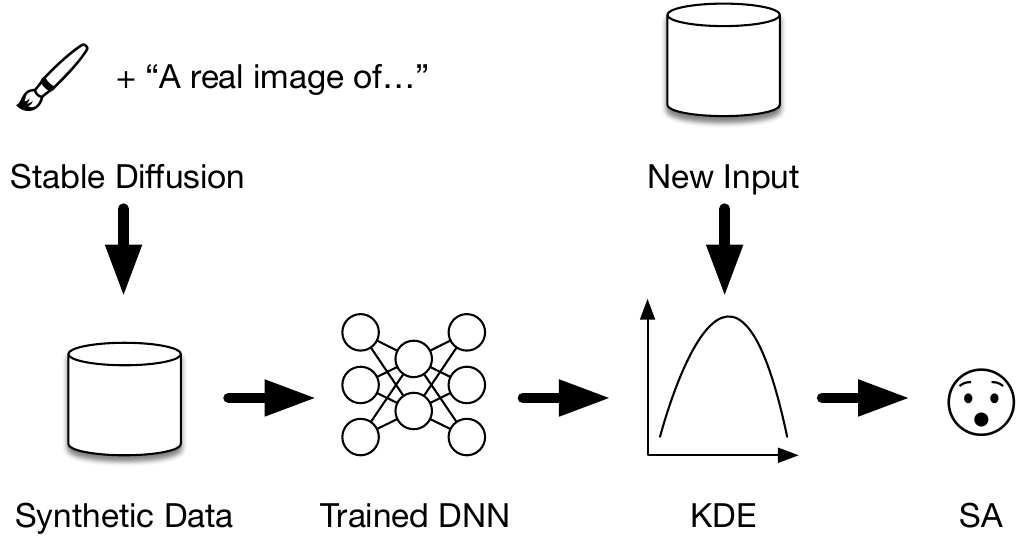}
    \caption{\name Surprise Adequacy Workflow\label{fig:dandi}}
    \end{subfigure}
    \caption{Overview of \name in comparison to original workflow of SA\label{fig:approach}}
\end{figure*}

\subsection{Stable Diffusion}
Stable Diffusion is a state-of-the-art text-to-image generative model capable of producing high-quality, diverse images guided by textual prompts~\cite{rombach2022high}. 
It transforms random noise into coherent images through diffusion modeling techniques~\cite{ho2020denoising},
specifically utilizing a latent diffusion model that operates within a compressed latent space.
Textual prompts are processed through an encoder to generate prompt embeddings, which guide the image generation process to align with the provided descriptions. 
The model employs a U-Net architecture~\cite{ronneberger2015u} augmented with a cross-attention mechanism~\cite{vaswani2017attention} to encode/decode images within this latent space.

In this context, the \textit{seed} is a critical parameter that initializes the random number generator used to produce the initial noise input for the diffusion process. Varying the seed alters the initial noise pattern, leading to different image outputs even when the same prompt is used. This capability allows us to generate a diverse set of images for each class label, as different seeds result in unique noise patterns that evolve into distinct images during the diffusion process. By leveraging different seeds, we enhance the diversity of the synthetic dataset and prevent duplicates.

\section{Approach: \name}
\label{sec:approach}

To overcome the dependency on the original training data for SA computation, we introduce \name, a technique that generates a surrogate input distribution using Stable Diffusion (Fig.~\ref{fig:dandi}). By creating a synthetic dataset that approximates the characteristics of the original training data, we can compute SA values without direct access to the original dataset.

Building on Stable Diffusion, \name generates a surrogate dataset by prompting the Stable Diffusion model with class labels from the target classification task. We use prompts in the format ``\textit{A real image of [label]},” replacing \textit{[label]} with each class name to ensure the generated images are relevant to the classification categories. 
To effectively represent the input distribution, we generate a diverse set of images per class, varying the random seed during image generation to enhance diversity and prevent duplicates.
Examples of the generated images are shown in Fig~\ref{fig:examples}, with 4 images per label.
Since Stable Diffusion operates optimally at a resolution of 512×512 pixels, we generate images at this size and downscale them as necessary to match the input requirements of the target DNNs.

With the surrogate dataset prepared, we compute the SA values for new inputs by measuring their dissimilarity to the activation patterns of the surrogate data. This involves extracting Activation Trace vectors from the DNN for both the surrogate dataset and the test inputs, and then calculating LSA based on these activations. By using the surrogate dataset generated through \name, we effectively approximate the SA values without the need for the original training data. This enables us to prioritize test inputs in scenarios where the training data is inaccessible, enhancing the applicability and efficiency of SA computation in practical settings.

\section{Experimental Settings}
\label{sec:rqs}
This section describes our RQs and experimental setup.

\subsection{Research Questions}
The goal of this study is to evaluate whether the surrogate input distribution generated by Stable Diffusion can effectively replace original training data in computing the out-of-distribution metric, SA, and to assess its effectiveness in test input prioritization for DNNs.

For a comprehensive evaluation, we perform the analysis in three ways: 1) compare the distributions of LSA values, 2) analyze the rank correlation between LSA values derived from both the original and synthetic datasets, and 3) assess the effectiveness of test input prioritization. These aspects are examined through the following research questions:

\subsubsection{RQ1. Comparison of SA Distributions}
How closely do the LSA value distributions generated by \name align with those derived from the original training dataset, indicating whether the underlying distribution properties are preserved?

\subsubsection{RQ2. Correlation Between SA Values}
To what extent do the LSA values obtained from \name mirror those from the original training dataset when ranking test inputs, thereby resulting in similar orderings?

\subsubsection{RQ3. Effectiveness in Test Input Prioritization}
How effectively does \name prioritize test inputs according to LSA value rankings, particularly in terms of test accuracy and mutation scores? To evaluate this, we compare the accuracy and mutation scores achieved when inputs are prioritized by \name against those achieved when prioritized using the original training dataset.

\subsection{Experimental Setup}
\subsubsection{Datasets and DL System} \
We conduct our experiments on two datasets: CIFAR-10~\cite{krizhevsky2014cifar} and ImageNet-1K~\cite{deng2009imagenet}, both of which are widely used in machine learning research for benchmarking image classification models.

\textbf{CIFAR-10} is a dataset consisting of 60,000 images divided into 10 different classes, with each image sized at \(32 \times 32\) pixels.
The dataset is split into 50,000 training images and 10,000 test images. For the neural network to use as the DNN under test, a 12-layer convolutional neural network with max-pooling and dropout layers are employed~\cite{Kim2019aa}. It was trained for 50 epochs to achieve 77.06\% accuracy on the test set.
 
\textbf{ImageNet-1K} (ILSVRC 2012 dataset) consists of 1.2 million training images and 50,000 validation images across 1,000 object categories, with images resized to \(224 \times 224\) pixels for classification models.
Due to its scale and diversity, it serves as a standard benchmark for evaluating deep learning models on large-scale image classification tasks~\cite{russakovsky2015imagenet}. 
To balance computational feasibility with representativeness, we select a subset of 15 categories from ImageNet by choosing five labels from each of three broad groups: food, animals and everyday items (Table \ref{tab:correlation_results}).
We evaluate our approach on a per-label basis, and utilize the validation and test datasets from ImageNet, providing approximately 120 images per label. 
For the food category, we incorporated the FoodNet101 dataset~\cite{bossard2014food}, which offers 1,000 samples per label, as we found an additional dataset suitable for this category.
Due to insufficient datasets, similar augmentation is not possible  for the other categories. 
For the DNN under test, we employ the pre-trained PyTorch implementation~\cite{paszke2019pytorch} of VGG16~\cite{simonyan2014very}, a convolutional neural network with 13 convolutional and three fully connected layers. The model achieves 71.59\% top-1 and 90.38\% top-5 accuracy on ImageNet.

\begin{figure}[b]
    \centering
    \includegraphics[width=0.78\columnwidth]{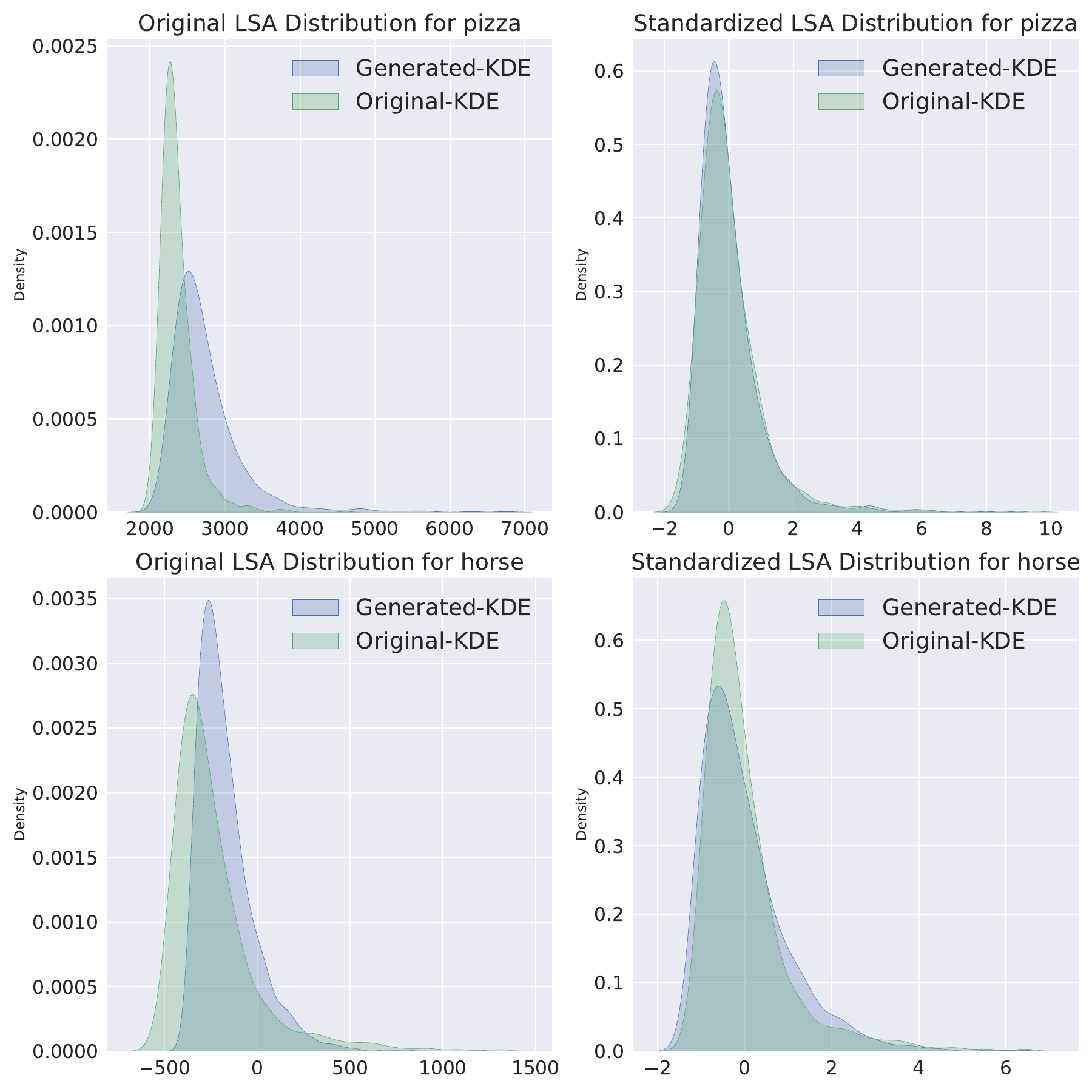}
    \caption{Distribution of LSA Scores and Standardized LSA Scores for pizza ImageNet-1K (top) and horse, CIFAR-10 (bottom)}
    \label{fig:rq1_dist}
\end{figure}

\begin{table}[t]
\caption{Average Test and Generated set Accuracy across different categories of ImageNet-1K and CIFAR10}
\centering
\begin{tabular*}{\columnwidth}{@{\extracolsep{\fill}} lrr}
\toprule
\textbf{Category} & \textbf{Test-set Acc (\%)} & \textbf{Generated-set Acc (\%)} \\
\midrule
\textbf{ImageNet-1K Food} & 79.05 & 91.88 \\
\textbf{ImageNet-1K Animals} & 91.56 & 99.10 \\
\textbf{ImageNet-1K Items} & 74.18 & 87.46 \\
\textbf{CIFAR10} & 77.06 & 82.60 \\
\bottomrule
\end{tabular*}
\label{tab:accuracy_results}
\end{table}

\subsubsection{Configurations}
For all research questions, LSA is computed using the activation traces from the penultimate layer, i.e., the input vector to the final neural network layer that produces the softmax logits, for both CIFAR-10 and ImageNet-1K datasets. 
Following the methodology suggested by Kim et al.~\cite{Kim2019aa}, we reduce the computational cost of Kernel Density Estimation (KDE) by excluding elements of the activation trace vectors with low variance. The bandwidth for KDE is selected according to Scott’s Rule to ensure appropriate smoothing.
To further facilitate the computation of KDE we perform Principal Component Analysis (PCA). This step is necessary because the aggregated activation traces tend to reside in a lower-dimensional subspace, resulting in a singular covariance matrix that the Gaussian KDE algorithm cannot process. By applying PCA for dimensionality reduction, we transform the data into a space with a non-singular covariance matrix, enabling KDE computation. Specifically, we reduce the dimensionality to 512 for the CIFAR-10 classifier and to 1,024 for VGG16.

\subsubsection{Generative Model}
For the generative model, we employ the pre-trained Stable Diffusion v1.4 provided by HuggingFace~\cite{rombach2022high}. This specific checkpoint was chosen for its demonstrated capability to generate photorealistic images from textual inputs. Following the authors' guidelines~\cite{rombach2022high}, we set the guidance scale to 7.5 and used 50 inference steps to generate the surrogate image dataset.

All experiments were performed on machines equipped with Intel i7-8700 CPUs and 32GB RAM GPU, running Ubuntu 20.04.6 LTS. CIFAR-10 and ImageNet-1K models are implemented using Torch v.2.0.1.

\section{Results}
\label{sec:results}
In this section, we present the results of our evaluation.

\subsection{Comparison of SA Distributions (RQ1)}
In this section, we address RQ1 by examining whether the LSA distributions generated by \name can serve as a surrogate for those derived from the original training dataset when calculating LSA. 

To assess the validity of the generated dataset, we analyze the top-1 accuracy achieved when using the generated data as test inputs for pre-trained models. The results for ImageNet-1K and CIFAR-10 are presented in Table \ref{tab:accuracy_results}. For both datasets, the generated dataset demonstrates a higher average accuracy than the test set, supporting its validity. Notably, we do not use these generated images directly for testing; rather, we employ their distribution as a surrogate for the training dataset in calculating LSA. This evaluation is conducted to confirm that the generated dataset is appropriate for this purpose and not merely a collection of random images. 

For descriptive analysis, we visualize and compare the LSA distributions using KDE plots. Due to space limitations, we present the results for a single label from ImageNet-1K and CIFAR-10 in Fig.~\ref{fig:rq1_dist}. Additional results can be found in our repository\footnote{\url{https://github.com/coinse/dandi}}.

Each result includes two KDE plots: one illustratingthe raw LSA score distributions and another showing the standardized distributions using z-score normalization. This normalization centers distributions around a mean of zero with unit variance, facilitating comparisons across different labels by eliminating scale differences. 
In the plots, the blue curve represents the LSA distribution from \name, while the green curve corresponds to the original dataset.
For both ImageNet-1K and CIFAR-10, the raw distributions often differ in range but generally exhibit a unimodal structure with similar patterns. After normalization, the alignment between distributions becomes even closer. 
Some distinctions persist; for instance, in CIFAR-10, the standardized distribution for the ``horse" label has a higher peak in the original dataset than in \name, indicating a greater density concentration. 
Despite these differences, the overall trends between the two distributions remain consistent post-normalization.

To further analyze the distributions, we compute the Jensen-Shannon (JS) divergence to quantify the differences between the LSA distributions generated by \name and those from the original datasets~\cite{lin1991divergence}.
The JS divergence ranges from 0 (identical distributions) to 1 (maximally different distributions). The complete results for all labels of ImageNet-1K and CIFAR-10 are presented in Table~\ref{tab:imagenet_distribution_analysis} and Table~\ref{tab:cifar10_distribution_analysis}. 

For the ImageNet-1K dataset, the JS divergence values are relatively low, ranging from 0.065 to 0.202 across different categories, with average values below 0.15. 
Similarly, for the CIFAR-10 dataset, the JS divergence values were generally low, averaging 0.131. However, certain labels such as ``truck'' exhibited higher values, up to 0.231, which is higher than any observed in the ImageNet-1K dataset. 
Upon closer examination, we discover that the CIFAR-10 dataset defines the ``truck" label to include only large trucks and explicitly excludes pickup trucks. 
Our general image generation prompt was ``A real image of a {truck}," which may have inadvertently resulted in images containing pickup trucks or other mismatched types, thereby causing discrepancies. 
To address this issue, we conducted an additional experiment using a more specific prompt for the ``truck" label: ``A real image of a big truck." This refinement reduced the JS divergence value from 0.235 to 0.217, illustrating that employing more precise prompts can enhance the alignment between generated images and the target dataset.

Our investigation into RQ1 indicates that for ImageNet-1K, the LSA distributions generated by \name closely align with those of the original training data after z-score normalization, as evidenced by low JS divergence values. 
In the case of CIFAR-10, despite some discrepancies attributable to factors like coarse prompts, the generated data still exhibited a high degree of similarity to the original dataset's LSA distributions.

\begin{tcolorbox}[colback=gray!20, boxsep=0pt, left=0.7mm, right=0.7mm, top=0.7mm, bottom=0.7mm]
\textbf{Answer to RQ1:} Our findings affirm that the distribution generated by \name effectively mirrors that of the original training dataset, validating its use as a surrogate in calculating LSA values.
\end{tcolorbox}

\begin{table}[t]
\caption{Jensen-Shannon Divergence (JSD): ImageNet-1K}
\centering
\begin{tabular}{lrlrlr}
\toprule
\multicolumn{2}{c}{\textbf{Food}} & \multicolumn{2}{c}{\textbf{Animals}} & \multicolumn{2}{c}{\textbf{Items}} \\
\midrule
\textbf{Label} & \textbf{JSD} & \textbf{Label} & \textbf{JSD} & \textbf{Label} & \textbf{JSD} \\
\midrule
pizza      & 0.086 & guinea pig & 0.169 & monitor    & 0.202 \\
ice cream  & 0.088 & hamster    & 0.117 & grandpiano & 0.124 \\
guacamole  & 0.121 & orangutan  & 0.164 & candle     & 0.132 \\
carbonara  & 0.098 & bee        & 0.136 & tripod     & 0.088 \\
burrito    & 0.065 & pelican    & 0.108 & binoculars & 0.098 \\
\midrule
\textbf{Average} & \textbf{0.092} & \textbf{Average} & \textbf{0.139} & \textbf{Average} & \textbf{0.129} \\
\bottomrule
\end{tabular}%
\label{tab:imagenet_distribution_analysis}
\end{table}

\begin{table}[t]
\caption{Jensen-Shannon Divergence (JSD): CIFAR-10}
\centering

\begin{tabular}{lrlr}
\toprule
\textbf{Label} & \textbf{JSD} & \textbf{Label} & \textbf{JSD} \\
\midrule
airplane   & 0.096 & frog       & 0.204 \\
automobile & 0.119 & dog        & 0.114 \\
bird       & 0.096 & horse      & 0.166 \\
cat        & 0.074 & ship       & 0.115 \\
deer       & 0.090 & truck      & 0.231 \\
\midrule
 & & \textbf{Average} & \textbf{0.131} \\
\bottomrule
\end{tabular}
\label{tab:cifar10_distribution_analysis}
\end{table}

\subsection{Correlation Between SA Values (RQ2)}
In this section, we address RQ2 by examining the rank correlation between the LSA values generated by \name and those derived from the original training dataset. 
We employ Spearman's rank-order correlation coefficient, $\rho$, to assess the relationship between the two sets of LSA values~\cite{spearman1987proof}: it is a non-parametric measure that assesses the monotonic relationship between two variables, ranging from \(-1\) (perfect negative correlation) to \(+1\) (perfect positive correlation), with \(0\) indicating no correlation. 

To determine the statistical significance of the observed correlation, we calculate the associated \emph{p}-value, representing the probability of obtaining such a correlation by chance under the null hypothesis of no correlation. A \emph{p}-value less than 0.05 indicates statistical significance. The correlation results for both the ImageNet-1K and CIFAR-10 datasets are presented in Table~\ref{tab:correlation_results}. We consider a correlation to be strong and significant when \(  \rho  > 0.7 \) and the \emph{p}-value is less than 0.05~\cite{dancey2007statistics}: results for all lables shown in Table~\ref{tab:correlation_results} are strong.

For ImageNet-1K, all Spearman correlation coefficients exceeded 0.7 with corresponding \emph{p}-values below 0.05, indicating strong positive correlations across these categories. 
Notably, the animals and items categories had smaller sample sizes, approximately 130 samples per label, which falls below the recommended minimum for reliable parametric significance testing. 
To address this, we employ permutation testing, a non-parametric method suitable for such conditions. Specifically, we shuffle the LSA values and recalculate Spearman correlation coefficients 10,000 times to construct an empirical null distribution. 
The results show that all labels within the animals and items categories yielded the minimal possible \emph{p}-value \((1/(n_{\text{permutations}} + 1) \approx 9.9 \times 10^{-5})\), confirming that the observed correlations are statistically significant and unlikely to have occurred by chance.

For CIFAR-10, the Spearman correlation coefficients for all labels are above 0.7, averaging 0.881, with corresponding \emph{p}-values all below 0.05, averaging \(3.67 \times 10^{-156}\).
These results align with those obtained from the ImageNet-1K dataset, indicating strong positive correlations across both datasets.

\begin{tcolorbox}[colback=gray!20, boxsep=0pt, left=0.7mm, right=0.7mm, top=0.7mm, bottom=0.7mm]
\textbf{Answer to RQ2:} The rank correlation analysis demonstrates a strong positive monotonic relationship between the LSA values derived from the original training dataset and those obtained using \name. The consistently high Spearman correlation coefficients and statistically significant \emph{p}-values indicate that the two sets of LSA values closely agree in prioritizing test inputs. 
\end{tcolorbox}

\begin{table}[t!]
\scriptsize
\centering
\caption{Correlation results for ImageNet-1K (Food, Animals, and Items) \& CIFAR-10\label{tab:correlation_results}}
\begin{tabular}{lrr|lrr}
\toprule
\multicolumn{3}{c|}{ImageNet-1K} & \multicolumn{3}{c}{CIFAR10} \\ \midrule
Label      & Size  & $\rho$ &Label         & Size  & $\rho$\\ \midrule
pizza      & 1,133 & 0.886  &airplane      & 1,000 & 0.949 \\
ice cream  & 1,079 & 0.922  &automobile    & 1,000 & 0.884 \\
guacamole  & 1,122 & 0.904  &bird          & 1,000 & 0.924 \\
carbonara  & 1,126 & 0.855  &cat           & 1,000 & 0.969 \\
burrito    & 1,109 & 0.872  &deer          & 1,000 & 0.948 \\\cmidrule{1-3}
Avg        & 1,114 & 0.888  &frog          & 1,000 & 0.712 \\\cmidrule{1-3}
guinea pig & 131   & 0.781  &dog           & 1,000 & 0.923 \\
hamster    & 149   & 0.828  &horse         & 1,000 & 0.801 \\
orangutan  & 150   & 0.738  &ship          & 1,000 & 0.918 \\
bee        & 139   & 0.910  &truck         & 1,000 & 0.783 \\
pelican    & 140   & 0.885  &bigtruck      & 1,000 & 0.820 \\\midrule
Avg        & 142   & 0.826  &Avg           & 1,000 & 0.881 \\\midrule
monitor    & 141   & 0.752  &              &       &       \\
grandpiano & 131   & 0.810  &              &       &       \\
candle     & 119   & 0.899  &              &       &       \\
tripod     & 105   & 0.919  &              &       &       \\
binoculars & 111   & 0.839  &              &       &       \\\cmidrule{1-3}
Avg        & 121   & 0.844  &              &       &       \\
\bottomrule
\end{tabular}
\end{table}
\begin{figure*}[t!]
    \centering
    \begin{subfigure}[t]{0.445\textwidth}
        \centering
        \includegraphics[width=\columnwidth]{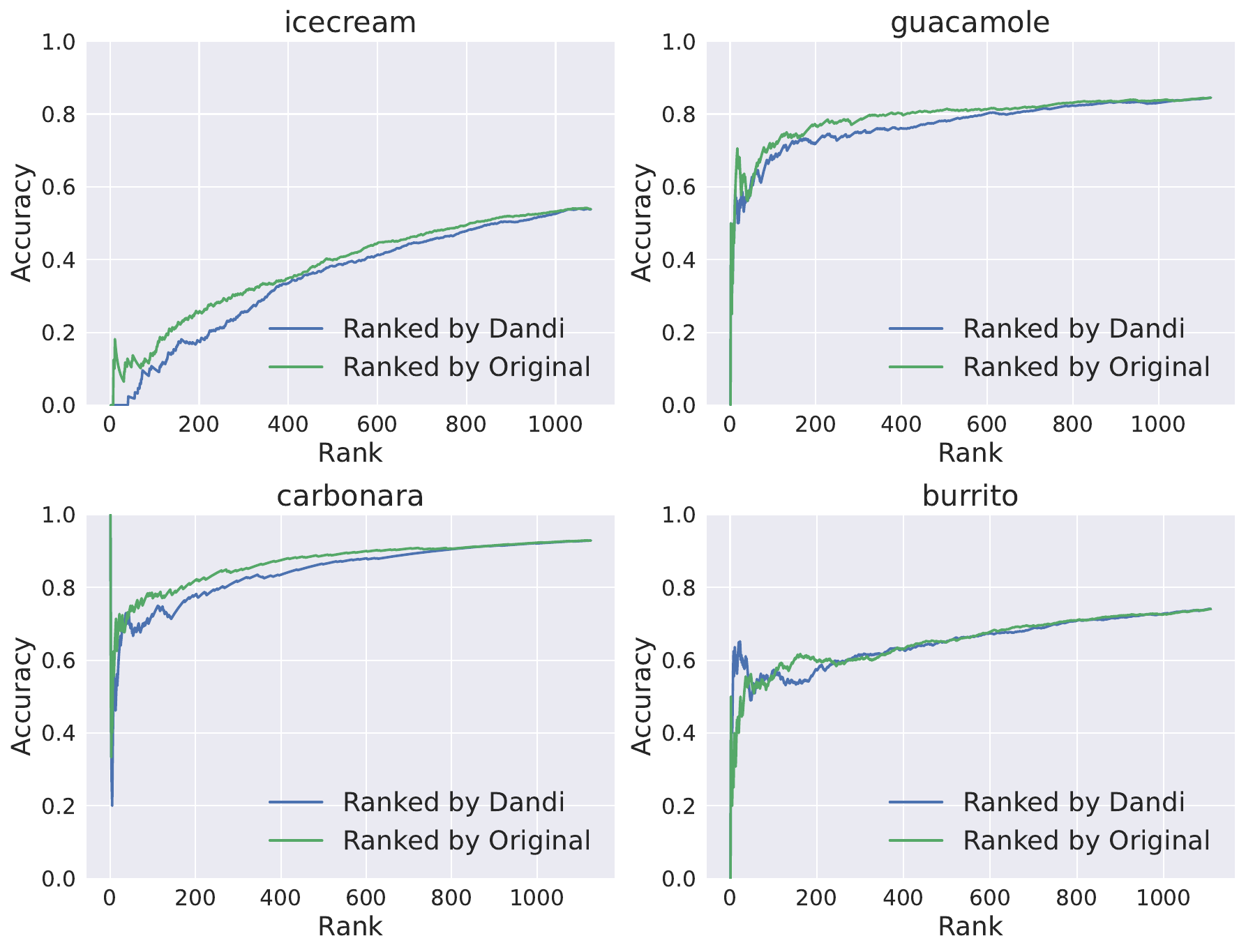}
        \caption{ImageNet-1K}
    \end{subfigure}
    \hfill
    \begin{subfigure}[t]{0.445\textwidth}
        \centering
        \includegraphics[width=\columnwidth]{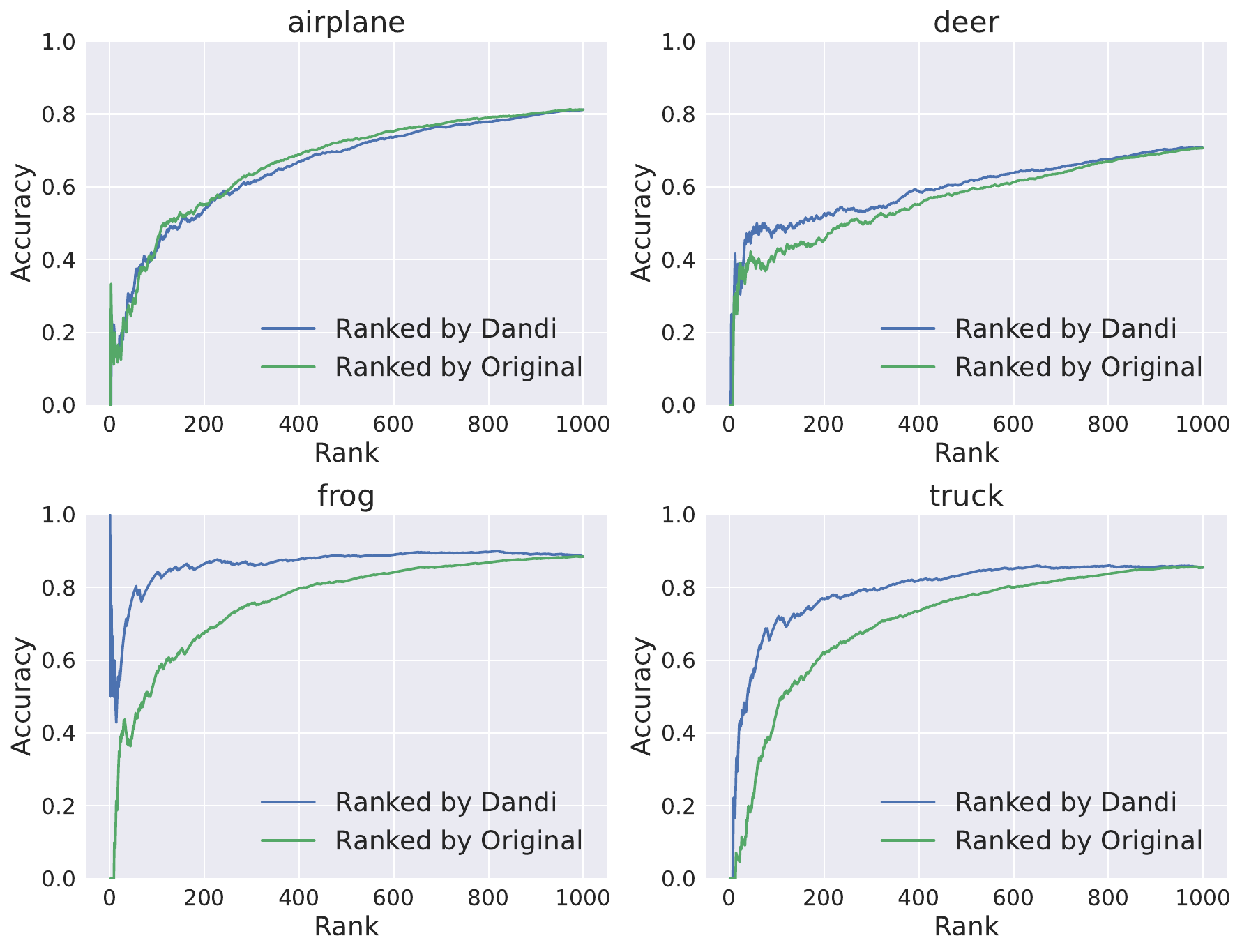}
        \caption{CIFAR-10}
    \end{subfigure}
    \caption{Impact of Input Prioritization on Model Accuracy}
    \label{fig:rq3_acc}
\end{figure*}

\subsection{Effectiveness in Test Input Prioritization (RQ3)}
In this section, we address RQ3 by examining input prioritization performance of \name and compare to those ranked using the original training dataset. 
We assess the effectiveness of \name by measuring test accuracy and analyzing mutant-killing capability based on LSA scores. 

To evaluate test accuracy, we sort the test inputs in descending order based on their LSA values, calculated using both the original training dataset and \name. 
For ImageNet, we focus exclusively on the food category. The test sets for the items and animals categories are relatively small (approximately 130 samples per label) and exhibited near-perfect accuracies, making it challenging to observe significant differences based on input prioritization.
For CIFAR-10, we evaluate all labels, as the test dataset contains a sufficient number of samples.

The results are presented in Fig.\ref{fig:rq3_acc}. Due to space constraints, for each dataset, we present the two labels with the highest and lowest correlation values: the top row displays the highest, and the bottom row shows the lowest.
In each graph, the green line represents accuracies achieved by prioritizing inputs using the original training dataset, while the blue line represents accuracies achieved by prioritizing inputs using \name. 

Both methods display similar trends: test accuracy increases with rank, corresponding to decreasing LSA values. This observation supports the findings outlined in the original surprise adequacy paper. 
We also observe that the strength of correlation is reflected in the accuracy measurements. For CIFAR-10, accuracy results closely align for labels, which had high correlation strength. 
For labels like ``frog'' and ``truck'' while there are some differences at the start of the rankings, both still exhibit the expected trend of increasing accuracy with higher ranks. 
For ImageNet-1K, accuracy trends are consistently aligned across all labels. The overall consistency between the two methods suggests that \name effectively approximates the original training dataset in prioritizing inputs.

To evaluate mutant-killing capability, we employ Gaussian Fuzzing (GF), a model-level mutation operator from DeepMutation, to generate mutant models. 
To ensure killable and non-trivial mutants, we adopt the binary search to tune the values of GF parameter (the ratio of the neurons affected by the mutation operator), instead of manually picking the parameter. 
We aim to discover the most challenging and yet killable configuration of the mutation operator. We adopt the statistical definition of killability provided by DeepCRIME, which involves using multiple instances of both the original and mutant models, as this approach offers more reliable results than relying on a single mutant instance.

For CIFAR-10, we train 20 independent instances of the original model and create mutants based on each one. However, for the ImageNet-1K, training 20 independent models is computationally intensive due to its scale and resource constraints. 
To address this, we simulate multiple instances by enabling dropout in the classifier part of the VGG16 model during inference, approximating the diversity of multiple models through stochastic outputs. 
pecifically, we perform 20 stochastic forward passes of the original model with dropout enabled and compare them to 20 stochastic forward passes of the mutant model derived from the original model. We assess killability for each class label based on these comparisons. 

Following the experimental setup of the original DeepMutation paper, we select inputs correctly classified by the original model and prioritize them in the descending order based on their LSA values. 
This approach targets inputs that are more surprising to the model, specifically those likely near the decision boundary, making them ideal for exposing discrepancies between the original and mutant models. 
Since mutants have perturbed decision boundaries, these prioritized inputs increase the likelihood of revealing misclassifications in the mutant models. 

We report the killability results using these inputs in Table \ref{tab:kill_results}. Each column displays the killability results using selected subsets of test data, comparing the original dataset with \name. 
For ImageNet-1K, we use subsets of 30, 50, and 70 samples; for CIFAR-10, we use subsets of 100, 300, and 500 samples from both datasets. 
Inputs selected from the original training set are labeled with '-O,' and those from \name are labeled with '-D.' For ImageNet-1K, using 30 samples, the original method killed mutants in 11 out of 15 class labels, while \name killed mutants in 12 out of 15. With 70 samples, the original method covered 14 class labels, whereas \name achieved kills in all 15. 
A similar trend is observed on CIFAR-10: with 100 samples, the original method killed mutants in 8 out of 10 class labels compared to 9 out of 10 for \name. At 500 samples, both methods killed mutants in all class labels. 
These results demonstrate that \name matches or surpasses the effectiveness of the original input selection method. It efficiently prioritizes inputs for testing DNN models, achieving higher mutant kill rates with fewer prioritized inputs.

\begin{tcolorbox}[colback=gray!20, boxsep=0pt, left=0.7mm, right=0.7mm, top=0.7mm, bottom=0.7mm]
\textbf{Answer to RQ3:} The effectiveness analysis shows that inputs prioritized by \name closely align with those ranked using the original training dataset. The test accuracy measurements and mutant-killing capability based on LSA scores indicate that \name effectively prioritizes test inputs.
\end{tcolorbox}

\begin{table}[t!]
\centering
\caption{Killability results: Killed labels are marked with checkmarks (\checkmark), non-killed labels are marked with dashes (-).\label{tab:kill_results}}
\begin{subtable}{\columnwidth}
\centering
\begin{tabular}{l c c c c c c}
\toprule
Label & 30-O & 30-D & 50-O & 50-D & 70-O & 70-D \\
\midrule
pizza          & \checkmark & \checkmark & \checkmark & \checkmark & \checkmark & \checkmark     \\
ice cream      & \checkmark & \checkmark & \checkmark & \checkmark & \checkmark & \checkmark     \\
guacamole      & \checkmark & \checkmark & \checkmark & \checkmark & \checkmark & \checkmark     \\
carbonara      & \checkmark & \checkmark & \checkmark & \checkmark & \checkmark & \checkmark     \\
burrito        & \checkmark & \checkmark & \checkmark & \checkmark & \checkmark & \checkmark     \\
\midrule
guinea pig & \checkmark & -          & \checkmark & -          & \checkmark & \checkmark \\
hamster    & -          & \checkmark & \checkmark & \checkmark & \checkmark & \checkmark \\
orangutan  & \checkmark & \checkmark & \checkmark & \checkmark & \checkmark & \checkmark \\
bee        & -          & -          & -          & \checkmark & \checkmark & \checkmark \\
pelican    & -          & -          & -          & \checkmark & -          & \checkmark \\
monitor    & \checkmark & \checkmark & \checkmark & \checkmark & \checkmark & \checkmark \\
\midrule
grandpiano & -          & \checkmark & \checkmark & \checkmark & \checkmark & \checkmark \\
candle     & \checkmark & \checkmark & \checkmark & \checkmark & \checkmark & \checkmark \\
tripod     & \checkmark & \checkmark & \checkmark & \checkmark & \checkmark & \checkmark \\
binoculars & \checkmark & \checkmark & \checkmark & \checkmark & \checkmark & \checkmark \\
\midrule
\textbf{Total} & 11/15      & 12/15      & 13/15      & 14/15      & 14/15      & \textbf{15/15} \\
\bottomrule
\end{tabular}
\caption{ImageNet-1K}
\end{subtable}

\vspace{5pt}
\begin{subtable}{\columnwidth}
\centering
\begin{tabular}{l c c c c c c}
\toprule
Label & 100-O & 100-D & 300-O & 300-D & 500-O & 500-D \\
\midrule
airplane   & \checkmark & \checkmark & \checkmark & \checkmark & \checkmark & \checkmark \\
automobile & \checkmark & \checkmark & \checkmark & \checkmark & \checkmark & \checkmark \\
bird       & \checkmark & \checkmark & \checkmark & \checkmark & \checkmark & \checkmark \\
cat        & -          & \checkmark & \checkmark & \checkmark & \checkmark & \checkmark \\
deer       & -          & -          & -          & -          & \checkmark & \checkmark \\
frog       & \checkmark & \checkmark & \checkmark & \checkmark & \checkmark & \checkmark \\
dog        & \checkmark & \checkmark & \checkmark & \checkmark & \checkmark & \checkmark \\
horse      & \checkmark & \checkmark & \checkmark & \checkmark & \checkmark & \checkmark \\
ship       & \checkmark & \checkmark & \checkmark & \checkmark & \checkmark & \checkmark \\
truck      & \checkmark & \checkmark & \checkmark & \checkmark & \checkmark & \checkmark \\
\midrule
\textbf{Total} & 8/10 & 9/10 & 9/10 & 9/10 & \textbf{10/10} & \textbf{10/10} \\
\bottomrule
\end{tabular}
\caption{CIFAR-10}
\end{subtable}
\end{table}

\section{Threats to Validity}
\label{sec:validity}

Threats to internal validity concern factors that may influence the conclusions drawn in this paper. 
In our study, these threats primarily involve the correctness of the implementation of the DL systems, the generative model, and the computation of SA values. 
To mitigate these risks, we either train classifier models using publicly available model architectures or use pre-trained models to ensure correct implementation; for the generative model, we exclusively use the publicly available pretrained Stable Diffusion model from Hugging Face. Our analyses were conducted using well-established statistical packages such as SciPy and scikit-learn.

Threats to external validity primarily concern the generalizability of our findings to other contexts.
In this study, we employed two DL systems and two datasets: CIFAR-10, representing a simpler dataset, and ImageNet, representing a more complex one. 
Due to computational constraints, our experiments on ImageNet are restricted to five labels within broad categories such as animals, food, and items. 
Future research should consider a more extensive set of labels.

Threats to construct validity concern whether our experimental setup accurately reflects the theoretical constructs we aim to study. 
In our approach, we simulate multiple model instances for ImageNet-1K by enabling dropout during inference in the VGG16 classifier instead of training 20 independent models due to the high cost of training. 
This method approximates model diversity through stochastic outputs; however, it may not fully capture the true variability of independently trained models with different initializations and training processes, potentially affecting the validity of our killability assessments for each class label.

\section{Discussion and Conclusion}
\label{sec:conclusion}
We introduce \name, a technique that leverages Stable Diffusion to generate surrogate input distributions for computing SA without requiring access to the original training data. 
By eliminating the dependence on proprietary or unavailable datasets, \name enhances the practicality of SA for testing DNNs. Our evaluation on classifiers trained on the CIFAR-10 and ImageNet-1K datasets demonstrates that SA values computed using synthetic data generated by \name highly correlate with those computed using the original data. 
This high correlation enables effective prioritization of inputs that reveal unexpected behaviors in DNN models. These findings indicate that \name improves the usability of SA for practical DNN testing.

Future work will consider applying \name to other data modalities, such as text, to broaden its applicability. 
Additionally, exploring its performance with various DNN architectures, including transformer-based networks, may provide additional insights. 
In summary, \name advances SA as a more accessible and practical metric for DNN testing by removing the need for original training data, thereby contributing to more effective testing practices in deep learning.

\bibliographystyle{acm}
\bibliography{newref}
\end{document}